\documentclass[twocolumn,showkeywords,preprintnumbers,amsmath,aps]{revtex4}
\usepackage{graphicx}
\usepackage{dcolumn}
\usepackage{bm}
\usepackage{color}
\begin{document}
\title{Signatures of non-adiabatic superconductivity in lithium-decorated graphene}
\author{Dominik Szcz{\c{e}}{\'s}niak$^{1}$}
\author{Rados{\l}aw Szcz{\c{e}}{\'s}niak$^{1, 2}$}
\affiliation{$^1$ Institute of Physics, Jan D{\l}ugosz University in Cz{\c{e}}stochowa, Ave. Armii Krajowej 13/15, 42-200 Cz{\c{e}}stochowa, Poland}
\affiliation{$^2$ Institute of Physics, Cz{\c{e}}stochowa University of Technology, Ave. Armii Krajowej 19, 42-200 Cz{\c{e}}stochowa, Poland}
\date{\today}
\begin{abstract}
Recent experimental studies confirmed that a long searched conventional superconducting phase in graphene can be induced via lithium deposition. However, the canonical character of the lithium-decorated graphene (LiC$_{6}$) superconductor postulated therein, as defined by the Bardeen-Cooper-Schrieffer (BCS) theory, is ambiguous due to the moderate electron-phonon coupling constant and low Fermi energy in this material. Herein, this issue is addressed within the Migdal-Eliashberg formalism and beyond it via the first-order electron-phonon vertex corrections, to account for the potentially pivotal effects, hitherto not captured. The conducted analysis yields similar value of the metal-superconductor transition temperature as in the previous studies ($T_{C} \sim 6$ K), yet for a weaker depairing electron correlations, which magnitude in the non-adiabatic regime approaches predictions of the Morel-Anderson model. Moreover, the characteristic ratio of the pairing gap and the inverse temperature is found to notably exceeds estimates of the BCS theory. In a results, it is argued that the superconducting state in LiC$_{6}$ may have non-canonical character strongly influenced by the non-adiabatic and retardation effects. Moreover, discussed phase appears as a vital case study, which suggests that other two-dimensional superconductors of hexagonal structure may exhibit similar non-canonical behavior as LiC$_{6}$.

\end{abstract}
\maketitle

\noindent {\bf Keywords:} phonon-mediated superconductivity, graphene-based superconductors, non-adiabatic and retardation effects

\vspace{0.5cm}

Since its discovery, the two-dimensional (2D) carbon allotrope known as graphene \cite{novoselov} established itself as an extraordinary solid-state platform for implementation of multiple electronic quantum phenomena \cite{castro}. However, a notable exception to these features is superconductivity, which, if possible, promises yet another breakthrough at the atomic scale associated with the substantial cross-disciplinary impact \cite{uchihashi, profeta, vafek}. The reason why the superconductivity is prevented in pristine graphene is the inherent nature of its charge carriers and the resulting small density of states around the Fermi level \cite{castro}. In this context, specific modifications of graphene are required to stabilize the superconducting phase.

Among multiple attempts to induce superconductivity in graphene \cite{cao, ludbrook, profeta, nandkishore, einenkel, heersche}, its decoration with lithium atoms appears as particularly promising one \cite{profeta, ludbrook}. This approach provides mature theoretical background \cite{vafek}, preliminary experimental verification \cite{ludbrook}, conventional character of the superconducting phase \cite{profeta}, and the well-established concept of using dopant atoms for promoting superconducting condensate \cite{weller}. In particular, the introduction of alkali adatoms to graphene breaks its chiral symmetry, generates new extended donor level in the system, and softens the undesirably energetic carbon out-of plane vibrations. Hereafter, the collective atomic-scale driven removal of the quantum confinement locates adatom state at the Fermi level, enhances the electron-phonon coupling constant ($\lambda$), and ultimately allows lithium-decorated graphene (LiC$_{6}$) to host superconducting phase.

Nonetheless, the fundamental understanding of superconducting phase in LiC$_{6}$ is still raising some questions, despite numerous studies in this field \cite{gholami, zheng, pesic, guzman, szczesniak, kaloni, profeta, ludbrook}. Of special attention is the experimental work by Ludbrook {\it et al.} \cite{ludbrook}; wherein a strong temperature ($T$) dependence is found for the pairing gap ($2\Delta(T)$, where $\Delta(T)$ is the order parameter) by using the angle resolved photoemission spectroscopy (ARPES). This study provides vital experimental verification of superconductivity in LiC$_{6}$, but presented therein canonical interpretation of this phase is ambiguous. In details, Ludbrook {\it et al.} preassume that in LiC$_{6}$ the characteristic ratio $2\Delta (0) / k_{B} T_{C}$ is equal to 3.5, where $k_{B}$ denotes the Boltzmann constant and $T_{C}$ is the transition temperature from normal to superconducting state. This is to say, the discussed experimental study suggests that LiC$_{6}$ is a canonical superconductor, according to the Bardeen-Cooper-Schrieffer (BCS) theory \cite{bardeen1, bardeen2}. However, this assumption is made by Ludbrook {\it et al.} only with the knowledge of the experimental $\Delta (T)$ value at 3.5 K and regardless of the electron-phonon coupling constant in LiC$_{6}$ ($\lambda \simeq 0.6$)  \cite{ludbrook, profeta, zheng} which exceeds the weak-coupling limit ($\lambda \simeq 0.5$) \cite{carbotte, cyrot}. In fact, this indicates potentially important role of the retardation effects not captured within the BCS theory \cite{szczesniak, carbotte}. Although, the canonical character of the LiC$_{6}$ superconductor is not substantially justified in \cite{ludbrook}, the aforementioned ratio is employed therein to calculate the $T_{C}$ value. Moreover, this critical temperature estimate is also used in the related theoretical study by Zheng and Margine \cite{zheng}, as a referential value for their investigations within the Eliashberg formalism \cite{eliashberg}. In a result, Zheng and Margine also suggest canonical character of the superconducting phase in LiC$_{6}$ \cite{zheng}. Furthermore, none of the discussed studies stress out that the relation $E_{F}>>\omega_{D}$ does not hold for the LiC$_{6}$, where $E_{F}$ stands for the Fermi energy and $\omega_{D}$ denotes Debye frequency. In fact, the $\omega_{D}/E_{F}$ ratio is equal to $\sim 0.145$ in LiC$_{6}$ and additionally implies the possible breakdown of the adiabatic approximation, according to the Migdal's theorem \cite{migdal}. This observation suggests need for the vertex corrections to the electron-phonon interaction \cite{pietronero}, \cite{grimaldi}, which are particularly important for superconducting materials with $\lambda < 1$ \cite{cai}, as in the case of LiC$_{6}$.

To address above ambiguities, herein the complementary analysis of the order parameter, the Coulomb pseudopotential, and the transition temperature in LiC$_{6}$ superconductor is provided within the Migdal-Eliashberg theory and beyond it. In particular, the Eliashberg equations are employed with \cite{pietronero}, \cite{grimaldi} and without \cite{eliashberg, migdal} the first-order vertex corrections, to properly account for the retardation and non-adiabatic effects in LiC$_{6}$ and to trace their influence on the calculated thermodynamic properties. Since occurrence of the non-adiabatic effects in LiC$_{6}$ is not certainly related to the breakdown of the Fermi liquid picture, vertex corrections are introduced within the perturbation scheme. Conventionally, the information about the electron-phonon processes in LiC$_{6}$ is encoded here within the concept of the Eliashberg spectral function $\alpha^{2}F\left(\omega\right)$, where $\alpha$ is the average electron-phonon coupling, $F\left(\omega\right)$ represents the phonon density of states and $\omega$ denotes the phonon frequency. In particular, calculations are conducted for the two representative and isotropically-averaged $\alpha^{2}F\left(\omega\right)$ functions, namely: the function obtained theoretically in \cite{profeta} by using the density functional theory, and the function derived directly from the ARPES experiment in \cite{ludbrook}. In this manner, it is possible to conveniently provide most complementary and comprehensive discussion of the LiC$_{6}$ thermodynamics, on the same footing with respect to the available experimental and theoretical predictions. For the completeness of the present analysis it is also important to note that the pairing gap in LiC$_{6}$ is actually predicted to have single anisotropic structure \cite{zheng, ludbrook}. Nonetheless, the isotropic approximation still provides averaged picture which is sufficient for the purpose of the present study, as shown later in the text. For more technical details on the computational methods please see Supplemental Material.

\begin{figure*}[ht!]
\includegraphics[width=\textwidth]{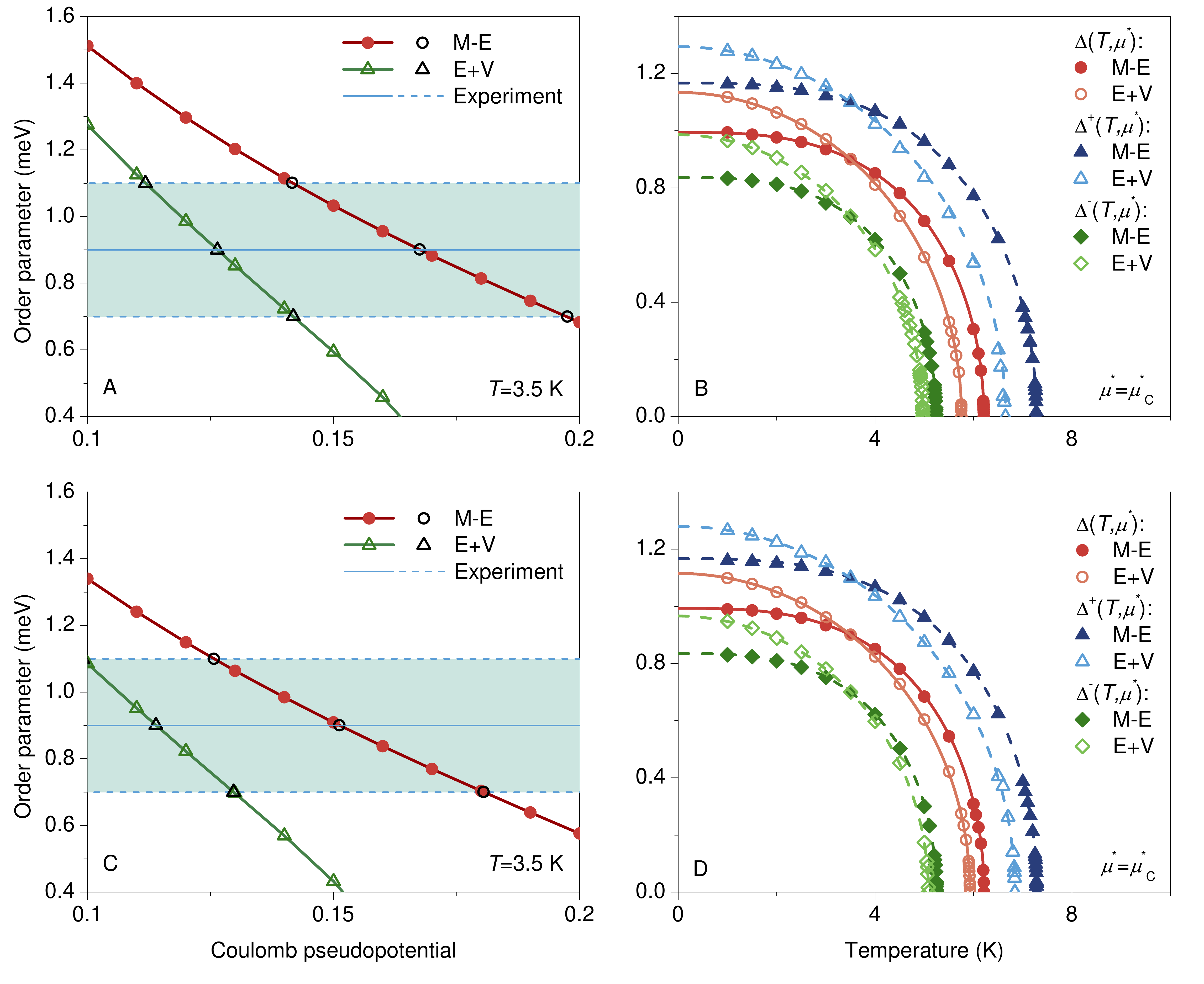}
\caption{The functional dependency of the order parameter in LiC$_{6}$ on the Coulomb pseudopotential ($\mu$) at 3.5 K ((A) and (C)) and the temperature ($T$) at the critical values of $\mu^{*}$ ((B) and (D)). Results obtained within the Eliashberg formalism with (E+V) and without (M-E) the first-order vertex corrections to the electron-phonon interaction are marked by the open and closed symbols, respectively. First row presents estimates obtained for theoretically calculated electron-phonon spectral function given in \cite{profeta}, whereas second row corresponds to the function experimentally derived in \cite{ludbrook}. For convenience, the referential experimental value of the order parameter at 3.5 K \cite{ludbrook} (solid blue line) along with its error bars (dashed blue lines) are depicted in (A) and (C). In (B) and (D), the corresponding order parameter ($\Delta(T, \mu^{*})$) values and their deviations ($\Delta^{+}(T, \mu^{*})$ and $\Delta^{-}(T, \mu^{*})$) are marked in by the solid and dashed lines, respectively.}
\label{fig01}
\end{figure*}

The numerical results obtained self-consistently within the in-house developed procedures \cite{szczesniak, durajski} are summarized in Fig. \ref{fig01}, which depicts the functional dependence of the order parameter on the Coulomb pseudopotential (first column of sub-figures) and temperature (second column of sub-figures), for the theoretically \cite{profeta} (first row of sub-figures) and experimentally (second row of sub-figures) \cite{ludbrook} derived $\alpha^{2}F\left(\omega\right)$ functions. In this context, the sub-figures \ref{fig01} (A) and (C) correspond to the initial computations, which are aimed at the determination of the physically-relevant critical value of the Coulomb pseudopotential ($\mu^{*}_{C}$) within the Eliashberg equations with (E+V) and without (M-E) vertex corrections. Note that this part of the analysis is carried out at T=3.5 K, to coincide with the temperature for which the value of the order parameter ($\Delta (3.5) \sim 0.9$ meV) was experimentally measured in \cite{ludbrook}. For convenience, in sub-figures \ref{fig01} (A) and (C) the $\Delta (3.5)$ value given in \cite{ludbrook} is marked by the vertical solid blue line, whereas its error bars are depicted as the vertical dashed lines above an below the referential value, to present the experimentally assessed region (shaded blue background color). In this manner, $\mu^{*}_{C}$ corresponds to the value for which the Eliashberg equations yields the $\mu^{*}$-dependent order parameter ($\Delta (3.5, \mu^{*})$) equal to the $\Delta (3.5)$ estimate given in \cite{ludbrook}. In other words, $\mu^{*}_{C}$ marks the point where the results of the Eliashberg equations cross the vertical blue line in sub-figures \ref{fig01} (A) and (C), as depicted therein by the open black circles (M-E results) and triangles (E+V results). In particular, the obtained $\mu^{*}_{C}$ values are equal to 0.126 (0.168) and 0.114 (0.142), as calculated within the Eliashberg equations with (without) vertex corrections for the theoretical \cite{profeta} and experimental \cite{ludbrook} $\alpha^{2}F\left(\omega\right)$ functions, respectively.

According to the obtained estimates, depicted in sub-figures \ref{fig01} (A) and (C), it can be qualitatively observed that both sets of results exhibit the same general behavior, which indicates that the inclusion of vertex corrections leads to the decrease of $\mu^{*}$ in the entire range of the $\Delta (T, \mu^{*})$ function, regardless of the employed Eliashberg function type. Naturally, results in sub-figures \ref{fig01} (A) and (C) differ from each other, what should be attributed to the differences in the shapes of the Eliashberg functions assumed for calculations. However, it can be argued that the $\mu^{*}$ values presented in sub-figure \ref{fig01} (C) are more truthful than results depicted in sub-figure \ref{fig01} (A), due to the fact that the former set is obtained on the basis of the Eliashberg function derived directly from the same experiment as the referential $\Delta (3.5)$ measure. In particular, among all the $\mu^{*}_{C}$ values presented here, the estimate provided by the E+V model for the experimental $\alpha^{2}F\left(\omega\right)$ function ($\mu^{*}_{C}=0.114$) is closest to the predictions of the Morel-Anderson model ($\mu^{*}_{C} \sim 0.1$) \cite{morel}. In fact, this estimate is the most physically relevant value of $\mu^{*}_{C}$, as expected from to the small difference between the electron-phonon coupling constant in LiC$_{6}$ ($\lambda \simeq 0.6$) and the upper weak-coupling limit for this observable ($\lambda \simeq 0.5$) \cite{carbotte, cyrot}. Further inspection allows to note that the $\mu^{*}_{C}$ value reported in the framework of the M-E model for the experimental Eliashberg function almost ideally corresponds to the value given by Zheng and Margine \cite{zheng} within the anisotropic Midgal-Eliashberg equations. This fact proves our former statement that the properties of LiC$_{6}$ calculated within the isotopic equations gives good approximation to the predictions of the anisotropic formalism. However, at the same time, results of the E+V model point out that both isotopic and anisotropic Migdal-Eliashberg equations somehow overestimate the magnitude of the electron-electron processes in LiC$_{6}$.

A notable changes due to the vertex corrections can be also noticed for the order parameter as a function of temperature for $\mu^{*}=\mu^{*}_{C}$, see sub-figures \ref{fig01} (B) and (D). In general, the vertex corrections cause decrease of the $T_{C}$ value and increase of the $\Delta(0, \mu^{*}_{C})$ level, for all considered cases. Interestingly, in the framework of the M-E formalism the $T_{C}$ estimates for both considered Eliashberg functions are the same and equal to 6.2 K. Likewise, the inclusion of the vertex corrections differentiate both values only slightly and gives 5.8 K and 5.9 K for the theoretically and experimentally derived $\alpha^{2}F\left(\omega\right)$ functions, respectively. In this context, the M-E estimates are higher than reported in \cite{zheng, ludbrook}, which equals to $\sim 5.8$ K, whereas the E+V calculations yields practically the same value as the one in the mentioned studies. The former disagreement is due to the fact that $\Delta (0, \mu^{*}_{C})$ is not pre-assumed here to be equal to the experimentally extracted $\Delta (3.5)$ value, as it was ambiguously done in \cite{zheng, ludbrook}. On the other hand, the latter convergence of the results is due to the interplay between the lack of the above pre-assumption and the inclusion of the vertex corrections. Note however, that although the discussed convergence occurs, only results presented here simultaneously give physically relevant $\mu^{*}_{C}$ value. Moreover, the effect of the vertex corrections is also visible when inspecting the $\Delta(0, \mu^{*}_{C})$ level, as mentioned above. In particular, already the M-E formalism proves here that LiC$_{6}$ is not a canonical example of the phonon-mediated superconductor, in contrary to suggestions made in \cite{zheng, ludbrook}. The $2\Delta (0, \mu^{*}_{C}) / k_{B} T_{C}$ ratio equals to 3.72 and 3.71 within the M-E model for the theoretical and experimental Eliashberg function, respectively. Therefore, presented estimates notably exceeds the canonical value of 3.5, which is predicted by the BCS theory. The discrepancy between the BCS predictions and results presented here is even more visible in the case when vertex corrections are included in the calculations. Specifically, the V+E model rises the $2\Delta (0, \mu^{*}_{C}) / k_{B} T_{C}$ ratio to the values of 4.57 and 4.37 for the theoretical and experimental Eliashberg functions, respectively.

In summary, the analysis of the superconducting state in LiC$_{6}$, as conducted here within the M-E and E+V models, reveals pivotal role of the non-adiabatic and retardation effects for the proper description of the discussed phase. The presented calculations clarify the observed interpretational ambiguities and provide complementary characteristic of the  LiC$_{6}$ superconductor. In particular, analyzed material is suggested to be a non-conventional phonon-mediated superconductor with a moderate Coulomb pseudopotential value and the critical temperature at the level of $\sim 6$ K. To this end, presented results shed a light on other 2D superconductors of hexagonal structure, which also present non negligible $\omega_{D}/E_{F}$ ratios (see Supplemental Material). In other words, present analysis can be considered as a case study of such materials, and a suggestion that superconducting state in hexagonal 2D superconductors is strongly non-adiabatic.

\begin{acknowledgments}
D. Szcz{\c e}{\' s}niak would like to acknowledge financial support of this work under Jan D{\l}ugosz University Research Grant for Young Scientists (grant no. DSM/WMP/5517/2017).
\end{acknowledgments}

\bibliographystyle{apsrev}
\bibliography{bibliography}
\end{document}